\documentclass[letterpaper, 12pt]{article}
\usepackage[english]{babel}
\usepackage{amsfonts,amsmath,amssymb,amsthm,latexsym,amscd,mathrsfs}
\usepackage{ifthen,cite}
\usepackage[bookmarksnumbered=true]{hyperref}
\usepackage{times,fancyhdr}

\usepackage{amsmath}
\numberwithin{equation}{section}

\usepackage{graphicx}

\hypersetup{pdfpagetransition={Split}}

\newcommand{\evenhead}{Author \ name}
\newcommand{\oddhead}{Article \ name}
\newcommand{\theArticleName}{Article \ name}

\newcommand{\FirstPageHeading}[1]{\thispagestyle{empty}%
\noindent\raisebox{0pt}[0pt][0pt]{\makebox[\textwidth]{\protect\footnotesize \sf}}\par}

\newcommand{\ArticleName}[1]{\renewcommand{\theArticleName}{#1}\vspace{-2mm}\par\noindent {\LARGE\bf  #1\par}}
\newcommand{\Author}[1]{\vspace{5mm}\par\noindent {\Large  #1\par} \par\vspace{2mm}\par}
\newcommand{\Address}[1]{\vspace{2mm}\par\noindent {\it #1} \par}
\newcommand{\Email}[1]{\ifthenelse{\equal{#1}{}}{}{\par\noindent {\rm E-mail: }{\it  #1} \par}}
\newcommand{\URLaddress}[1]{\ifthenelse{\equal{#1}{}}{}{\par\noindent {\rm URL: }{\tt  #1} \par}}
\newcommand{\EmailD}[1]{\ifthenelse{\equal{#1}{}}{}{\par\noindent {$\phantom{\dag}$~\rm E-mail: }{\it  #1} \par}}
\newcommand{\URLaddressD}[1]{\ifthenelse{\equal{#1}{}}{}{\par\noindent {$\phantom{\dag}$~\rm URL: }{\tt  #1} \par}}

\newcommand{\Abstract}[1]{\vspace{6mm}\par\noindent\hspace*{10mm}
\parbox{140mm}{\small {\bf Abstract.} #1}\par}
\newcommand{\Keywords}[1]{\vspace{3mm}\par\noindent\hspace*{10mm}
\parbox{140mm}{\small {\bf Key words:} \rm #1}\par}
\newcommand{\Classification}[1]{\vspace{3mm}\par\noindent\hspace*{10mm}
\parbox{140mm}{\small {\it 2020 Mathematics Subject Classification:} \rm #1}\vspace{3mm}\par}
\newcommand{\ShortArticleName}[1]{\renewcommand{\oddhead}{#1}}
\newcommand{\AuthorNameForHeading}[1]{\renewcommand{\evenhead}{#1}}
\long\def\@makecaption#1#2{
  \sbox\@tempboxa{\small \textbf{#1.}\ \ #2}%
  \ifdim \wd\@tempboxa >\hsize
    {\small \textbf{#1.}\ \ #2}\par \else
    \global \@minipagefalse
    \hb@xt@\hsize{\hfil\box\@tempboxa\hfil}%
  \fi \vskip\belowcaptionskip}

\def\numberwithin#1#2{\@ifundefined{c@#1}{\@nocounterr{#1}}{%
  \@ifundefined{c@#2}{\@nocnterr{#2}}{%
  \@addtoreset{#1}{#2}%
  \toks@\@xp\@xp\@xp{\csname the#1\endcsname}%
  \@xp\xdef\csname the#1\endcsname
    {\@xp\@nx\csname the#2\endcsname.\the\toks@}}}}
\def\E^#1{{\buildrel #1 \over\vee}}
\newtheorem{theorem}{Theorem}
\newtheorem{criterion}{Criterion}

{\theoremstyle{definition}

}

\usepackage{tikz}

\begin{document}

\FirstPageHeading{V.I. Gerasimenko}

\ShortArticleName{Cumulant expansions}

\AuthorNameForHeading{V.I. Gerasimenko}

\ArticleName{\textcolor{blue!50!black}{Cluster expansions of particle system state \\ with topological nearest-neighbor interaction}}

\Author{V.I. Gerasimenko$^\ast$\footnote{E-mail: \emph{gerasym@imath.kiev.ua}} and
I.V. Gapyak$^{\ast,}$$^\ast$$^\ast$\footnote{E-mail: \emph{gapjak@ukr.net}}}

\Address{$^\ast$\hspace*{1mm}Institute of Mathematics of NAS of Ukra\"{\i}ne,\\
    \hspace*{3mm}3, Tereshchenkivs'ka Str.,\\
    \hspace*{3mm}01601, Ky\"{\i}v-4, Ukra\"{\i}ne}

\Address{$^\ast$$^\ast$Taras Shevchenko National University of Ky\"{\i}v,\\
    \hspace*{3mm}Department of Mechanics and Mathematics,\\
    \hspace*{3mm}2, Academician Glushkov Av.,\\
    \hspace*{3mm}03187, Ky\"{\i}v, Ukra\"{\i}ne}

\bigskip

\Abstract{The article presents a concept of a cumulant representation for distribution functions
describing the states of many-particle systems with topological nearest-neighbor interaction.
A solution to the Cauchy problem for the hierarchy of nonlinear evolution equations for the
cumulants of distribution functions of such systems is constructed. The connection between the
constructed solution and the series expansion structure for a solution to the Cauchy problem of
the BBGKY hierarchy has been established. Furthermore, the expansion structure for a solution to
the Cauchy problem of the hierarchy of evolution equations for reduced observables of topologically
interacting particles is established.
}

\bigskip


\vspace{2pc}
\Classification{35C10; 35Q20; 82C05; 82D05; 70F45}

\makeatletter
\renewcommand{\@evenhead}{
\hspace*{-3pt}\raisebox{-7pt}[\headheight][0pt]{\vbox{\hbox to \textwidth {\thepage \hfil \evenhead}\vskip4pt \hrule}}}
\renewcommand{\@oddhead}{
\hspace*{-3pt}\raisebox{-7pt}[\headheight][0pt]{\vbox{\hbox to \textwidth {\oddhead \hfil \thepage}\vskip4pt\hrule}}}
\renewcommand{\@evenfoot}{}
\renewcommand{\@oddfoot}{}
\makeatother

\newpage
\vphantom{math}

\protect\textcolor{blue!50!black}{\tableofcontents}

\vspace{0.8cm}

\textcolor{blue!50!black}{\section{Introduction}}
As is known \cite{CGP97},\cite{GG21}, the evolution of the state of a many-particle system is traditionally
described within the framework of a probability distribution function, which is governed by the Liouville
equation. In the paper \cite{GG22}, an alternative approach to describing the evolution of the state was
developed, which consists of using functions defined as cumulants of probability distribution functions.
Cumulants of probability distribution functions are interpreted as correlations of particle states,
therefore the term correlation functions \cite{G56,P62,B71} is also used for them. The evolution of
correlation functions is described by a hierarchy of nonlinear Liouville evolution equations, constructed
for a system of many hard spheres in the paper \cite{GG22}.

A few years ago, to describe complex systems, particularly systems of mathematical biology, "topological"
interaction models were introduced \cite{BD16,BD17,DP19,BPR25}. Namely, the notion of interaction was
introduced, which is described as a constituent of the system reacting to the presence of another not
according to the distance between them, analogous to many-particle systems, but according to the rank of
proximity or their order of preference. Interactions between living beings in nature are weighted as a
function of their rank, regardless of relative distance, that is, the probability of an individual interacting
with its nearest neighbor is the same whether that individual is close or far away. This new type of interaction
was called "topological" in contrast to the usual "metric" interaction, a function of the relative distance
between entities. A typical sample of a topological interaction model in many-particle systems is a one-dimensional
system of particles with the interaction between nearest neighbors, particularly the well-known Toda lattice
which is a model for a crystal in solid-state physics, or a one-dimensional non-symmetric system of hard rods,
which is described by functions not invariant under the particle renumbering \cite{G1},\cite{PG83}.

The article formulates the concept of a cumulant representation for distribution functions, which describes
the evolution of state correlations of many-particle systems with topological interaction. It is established
that the structure of cumulant expansions for such systems depends on the symmetry of the probability distribution
functions of many-particle systems.

In the space of sequences of integrated functions, a solution to the Cauchy problem for the hierarchy of evolution
nonlinear equations for the cumulants of the distribution functions of such systems is constructed. Based on the
dynamics of correlations, expansions into series were determined for the reduced distribution functions and their
cumulants known as the reduced correlation functions. In particular, it enabled us to explain the structure of
the generating operators of the non-perturbative solution of the Cauchy problem to the BBGKY hierarchy
(Bogolyubov--Born--Green--Kirkwood--Yvon) for particle systems with topological nearest-neighbor interaction.
The appendix also establishes the expansion structure of a non-perturbative solution to the Cauchy problem of the
hierarchy of evolution equations for reduced observables of topologically interacting particles.

\textcolor{blue!50!black}{\section{Topological nearest-neighbor interaction of many particles}}
In one-dimensional space, we consider a system of many identical particles that interact with their nearest
neighbors through a short-range interaction potential $\Phi$, namely, $\Phi(q)=0,$ if $|q|>R>0$, with a hard
core of length $\sigma$, i.e. $\Phi(\sigma)=+\infty$. Let the interaction potential $\Phi$ satisfy the following
conditions: $\Phi\in C^2[\sigma,R]$. For the configurations of many particles with such "topological" nearest-neighbor
interaction, the following inequalities hold: $\sigma+q_i\leq q_{i+1}$, where $q_i\in\mathbb{R}$ is the coordinate
of the center of the $i$-th particle. Since a system of a non-fixed number of ordered particles is considered, it
natural to number the particles using positive and negative integers from the set $\mathbb{Z}\setminus \{0\}$ as
follows \cite{G1},\cite{PG83}. The phase coordinates of a single-particle subsystem are numbered as canonical
quantities characterizing the first particle. The phase coordinates of particles in many-particle subsystems located
to the right of the first particle are numbered with positive integers, and those located to the left of the first
particle are numbered with negative integers.

Let $L_{n_1+n_2}^1$ be the Banach space of double sequences $f=\{f_{n_1+n_2}(x_{-n_1},\ldots,x_{n_2})\}_{n_1+n_2\geq0}$
of integrable functions $f_{n_1+n_2}(x_{-n_1},\ldots,x_{n_2})$ defined on the phase space
$\mathbb{R}^{{n_1+n_2}}\times(\mathbb{R}^{{n_1+n_2}}\setminus \mathbb{W}_{n_1+n_2})$, equal zero on the set of
forbidden configurations $\mathbb{W}_{n_1+n_2}\equiv \{(q_{-n_1},\ldots,q_{-1},q_{1},\ldots,q_{n_2})\in \mathbb{R}^{n_1+n_2}\mid\sigma+q_i<q_{i+1}\,\mathrm{at}\,\mathrm{least}\,\mathrm{for}\,\mathrm{one}\,
\mathrm{pair}\,(i,i+1)\in ((-n_1,-n_1 +1),\ldots,(-1,1),\ldots,(n_2-1, n_2))\}$, and non-symmetric with
respect to the permutation of arguments $x_i\equiv(q_i,p_i)\in \mathbb{R}\times\mathbb{R}$ with the norm
\begin{eqnarray*}
   &&\|f_{n_1+n_2}\|_{L_{n_1+n_2}^1}=\int_{\mathbb{R}^{n_1}\times\mathbb{R}^{n_2 }}
    dx_{-n_1}\ldots dx_{n_2}\,|f_{n_1+n_2}(x_{-n_1},\ldots,x_{n_2})|.
\end{eqnarray*}

Under the conditions formulated above for the interaction potential $\Phi$, the functions
$X_{i}(-t),\,i\in(-n_1,\ldots,-1,1,\ldots,n_2)$ are global-in-time solutions of the Cauchy problem
to the Hamilton equations of the system $n_1+n_2$ of particles with initial data
$X_{i}(0)=x_i,\,i\in(-n_1,\ldots,-1,1,\ldots,n_2)$ \cite{CGP97}. Then in the space of integrable functions
$L_{n_1+n_2}^1$, the group of operators of the Liouville equation is defined as follows \cite{CGP97},\cite{BL}:
\begin{eqnarray}\label{S*}
   &&\hskip-7mm S_{n_1+n_2}^\ast(t)f_{n_1+n_2}\equiv(S_{n_1+n_2}^\ast(t,-n_1,\ldots,n_2)\,
         f_{n_1+n_2})(x_{-n_1},\ldots,x_{n_2})\doteq\\
   &&\begin{cases}
         f_{n_1+n_2}(X_{-n_1}(-t,x_{-n_1},\ldots,x_{n_2}),\ldots,X_{n_2}(-t,x_{-n_1},\ldots,x_{n_2})),\\
         \hskip+45mm(x_{-n_1},\ldots,x_{n_2})\in(\mathbb{R}^{n_1+n_2}\times(\mathbb{R}^{n_1+n_2}\setminus
                    \mathbb{W}_{n_1+n_2})),\\
         0, \hskip+41mm(q_{-n_1},\ldots,q_{n_2})\in\mathbb{W}_{n_1+n_2}.
\end{cases}
\nonumber
\end{eqnarray}
In the space $L_{n_1+n_2}^1$, the one-parameter mapping \eqref{S*} is a strongly continuous group of isometric
operators. The infinitesimal generator of the group of operators \eqref{S*} has the following structure
\begin{eqnarray}\label{Lstar}
       \mathcal{L}^{\ast}_{n_1+n_2}=\sum_{j\in(-n_1,\ldots,-1,1,\ldots,n_2)}\mathcal{L}^{\ast}(j)+
        \sum_{(j,j+1)\in ((-n_{1},-n_{1}+1),\ldots,(n_{2}-1,n_{2}))}\mathcal{L}^{\ast}_{\mathrm{int}}(j,j+1),
\end{eqnarray}
where on the subspace of continuously differentiable functions with compact supports the Liouville operators
are determined by the following formulas, respectively \cite{PG83}:
\begin{eqnarray*}
    &&\hskip-12mm(\mathcal{L}^{\ast}(j)f_{n_1+n_2})(x_{-n_1},\ldots,x_{n_2})\doteq
        - p_j\frac{\partial}{\partial q_j}f_{n_1+n_2}(x_{-n_1},\ldots,x_{n_2}),\\
    &&\hskip-12mm(\mathcal{L}_{\mathrm{int}}^{\ast}(j,j+1)f_{n_1+n_2})(x_{-n_1},\ldots,x_{n_2})\doteq
    \frac{\partial}{\partial q_{j}}\Phi(q_{j}-q_{j+1})
   (\frac{\partial}{\partial p_{j}}-\frac{\partial}{\partial p_{j+1}})f_{n_1+n_2}(x_{-n_1},\ldots,x_{n_2}).
\end{eqnarray*}

Let $D^0_{n_1+n_2}\in L_{n_1+n_2}^1(\mathbb{R}^{n_1+n_2}\times(\mathbb{R}^{n_1+n_2}\setminus\mathbb{W}_{n_1+n_2}))$
be a probability distribution function. Then all possible states of the system of $n_1+n_2$ particles at an arbitrary
instant $t\in\mathbb{R}$ are described by the probability distribution function
\begin{eqnarray}\label{D}
    &&D_{n_1+n_2}(t)=S_{n_1+n_2}^\ast(t)D^0_{n_1+n_2},
\end{eqnarray}
which is a unique solution to the Cauchy problem for the Liouville equation.
This is a strong solution for initial data from the subspace of finite sequences of continuously differentiable
functions with compact supports. For arbitrary initial data from the space $L^{1}_{n_1+n_2},\,n_1+n_2\geq1$, it is
a weak solution \cite{CGP97}.

\textcolor{blue!50!black}{\section{Cluster and cumulant expansions of distribution functions}}
For a system of a non-fixed number of particles \cite{PG83}, which will be considered further, the state is
described by a double sequence $D=\{D_{n_1+n_2}(x_{-n_1},\ldots,x_{-1},x_{1},\ldots,x_{n_2})\}_{n_1+n_2 \geq 0}$
of non-symmetric to the permutation of their arguments $x_i\equiv(q_i,p_i)\in \mathbb{R}\times\mathbb{R}$
probability distribution functions, which are defined on the phase space
$\mathbb{R}^{n_1+n_2}\times(\mathbb{R}^{n_1+n_2}\setminus \mathbb{W}_{n_1+n_2})$.

Let us introduce a sequence of correlation functions $g=\{g_{s}(x_{-s_1},\ldots,x_{s_2})\}_{s=s_1+s_2\geq 0}$
of particle states as cumulants (semi-invariants) of probability distribution functions using cluster expansions:
\begin{eqnarray}\label{cexp}
   &&\hskip-12mm D_{s_1+s_2}(x_{-s_1},\ldots,x_{s_2})=
       \sum\limits_{\mbox{\scriptsize $\begin{array}{c}\mathrm{P}:(x_{-s_1},\ldots,x_{s_2})=
       \bigcup_{i}X_{i}\end{array}$}}\prod_{X_i\subset\mathrm{P}}g_{|X_i|}(X_i),
       \quad s_1+s_2\geq1,
\end{eqnarray}
where the symbol $\sum_\mathrm{P}$ denotes the sum over all ordered partitions $\mathrm{P}$ of the partially
ordered set $(x_{-s_1},\ldots,$ $x_{-1},x_{1},\ldots,x_{s_2})$ into $|\mathrm{P}|$ nonempty partially ordered
subsets $X_i\in(x_{-s_1},\ldots,x_{-1},x_{1},\ldots,$ $x_{s_2})$ that do not intersect. The origin of the
presented approach of describing states in terms of correlation functions will be justified in Section 5.
We give examples of cluster expansions \eqref{cexp}:
\begin{eqnarray*}
   &&\hskip-12mm D_{0+1}(x_1)=g_{0+1}(x_1), \\
   &&\hskip-12mm D_{1+1}(x_{-1},x_1)=g_{1+1}(x_{-1},x_1)+g_{1+0}(x_{-1})g_{0+1}(x_1),\\
   &&\hskip-12mm D_{1+2}(x_{-1},x_1,x_2)=g_{1+2}(x_{-1},x_1,x_2)+g_{1+1}(x_{-1},x_1)g_{0+1}(x_2)+\\
   && g_{1+0}(x_{-1})g_{0+2}(x_1,x_2)+g_{1+0}(x_{-1})g_{0+1}(x_1)g_{0+1}(x_2), \\
   &&\vdots
\end{eqnarray*}

It should be noted that the concept of cluster expansions was initially introduced for Gibbs distribution
functions in the work \cite{MM41}. This was to describe the equilibrium state using Ursell functions for
many-particle systems characterized by distribution functions symmetric concerning permutations of arguments.
It is important to highlight that the structure of cluster expansions \eqref{cexp} differs significantly
from the structure of analogous cluster expansions for systems describing by symmetric functions \cite{GG22}.

On the set $\mathbb{R}^{s_1+s_2}\setminus\mathbb{W}_{s_1+s_2}$ of allowed configurations, the solutions
of the recursive relations \eqref{cexp} are represented by the following cumulant expansions:
\begin{eqnarray}\label{cuexp}
   &&\hskip-12mm g_{s_1+s_2}(x_{-s_1},\ldots,x_{s_2})=
        \sum\limits_{\mbox{\scriptsize $\begin{array}{c}\mathrm{P}:(x_{-s_1},\ldots,x_{s_2})=
       \bigcup_{i}X_{i}\end{array}$}}(-1)^{|\mathrm{P}|-1}\,
       \prod_{X_i\subset \mathrm{P}}D_{|X_i|}(X_i), \quad s_1+s_2\geq1,
\end{eqnarray}
where the used denotations are similar to that of the recursive equations \eqref{cexp}. We give examples
of cumulant expansions \eqref{cuexp}:
\begin{eqnarray*}
   &&\hskip-12mm g_{0+1}(x_1)=D_{0+1}(x_1), \\
   &&\hskip-12mm g_{1+1}(x_{-1},x_1)=D_{1+1}(x_{-1},x_1)-D_{1+0}(x_{-1})D_{0+1}(x_1),\\
   &&\hskip-12mm g_{1+2}(x_{-1},x_1,x_2)=D_{1+2}(x_{-1},x_1,x_2)-D_{1+1}(x_{-1},x_1)D_{0+1}(x_2)-\\
   &&D_{1+0}(x_{-1})D_{0+2}(x_1,x_2)+D_{1+0}(x_{-1})D_{0+1}(x_1)D_{0+1}(x_2),\\
   &&\vdots
\end{eqnarray*}

Thus, the structure of the expansions \eqref{cuexp} for correlation functions is such that they have
the meaning of cumulants (semi-invariants) of the corresponding order of probability distribution
functions \cite{GG21},\cite{D19},\cite{DS}. We note that correlation functions provide an equivalent
method of describing the states of many-particle systems analogous to distribution functions. The
interpretation of correlation functions is related to the fact that the state of statistically independent
particles on allowed configurations is described by the product of one-particle correlation functions,
i.e., the probability distribution functions of each particle.

To compare the structure of the expansions \eqref{cexp} and \eqref{cuexp} with similar expansions
for symmetric functions, we will represent them on sequences of functions. In the set of double
sequences $f=\{f_{n_1+n_2}(x_{-n_1},\ldots,x_{-1},x_{1},\ldots,x_{n_2})\}_{n_1+n_1=n\geq0}$ of
measurable functions $f_{n_1+n_2}$, where $f_0$ is arbitrary number, we introduce the following
tensor $\star$-product
\begin{eqnarray*}
  &&(f_1\star f_2)_{|X|}(X)\doteq\sum_{Y\subset X}(f_1)_{|Y|}(Y)\,(f_2)_{|X \setminus Y|}(X\setminus Y),
\end{eqnarray*}
where the symbol $\sum_{Y\subset X}$ denotes the sum over all possible partially ordered subsets $Y$ of
the partially ordered set $X\equiv(x_{-n_1},\ldots,x_{-1},x_{1},\ldots,x_{n_2})$.

On the set of double sequences $f$ of measurable functions, the mapping $(I-\circ)^{-{\mathbb I}_{\star}}$
is defined as the $\star$-resolvent, namely,
\begin{eqnarray}\label{cl}
  &&\hskip-5mm(I-f)^{-{\mathbb I}_{\star}}=I+\sum\limits_{n=1}^{\infty}f^{\star n},
\end{eqnarray}
where $I=(1,0,0,\ldots)$ is a unit sequence, and is represented component by component form by the following
expansions:
\begin{eqnarray*}
  &&\hskip-5mm\big((I-f)^{-{\mathbb I}_{\star}}\big)_{s_1+s_1}(x_{-s_1},\ldots,x_{s_2})=\delta_{s_1+s_2,0}+
       \sum\limits_{\mbox{\scriptsize $\begin{array}{c}\mathrm{P}:(x_{-s_1},\ldots,x_{s_2})=
       \bigcup_{i}X_{i}\end{array}$}}\prod_{X_i\subset\mathrm{P}}f_{|X_i|}(X_i),\\
    &&\hskip-5mm  s_1+s_2\geq1.
\end{eqnarray*}
The inverse mapping $I-(I+\circ )^{-{\mathbb I}_{\star}}$ to the mapping $(I-\circ )^{-{\mathbb I}_{\star}}$
is defined in terms of the $\star$-product by the following series
\begin{eqnarray}\label{cu}
  &&\hskip-5mm  I-(I+f)^{-{\mathbb I}_{\star}}=\sum_{n=1}^{\infty}(-1)^{n-1}f^{\star n},
\end{eqnarray}
which is represented in component-wise form by the following expansions:
\begin{eqnarray*}
  &&\hskip-5mm\big(I-(I+f)^{-{\mathbb I}_{\star}}\big)_{s_1+s_1}(x_{-s_1},\ldots,x_{s_2})=
       \sum\limits_{\mbox{\scriptsize $\begin{array}{c}\mathrm{P}:(x_{-s_1},\ldots,x_{s_2})=
       \bigcup_{i}X_{i}\end{array}$}}(-1)^{|\mathrm{P}|-1}\,\prod_{X_i\subset \mathrm{P}}f_{|X_i|}(X_i),\\
    &&\hskip-5mm  s_1+s_2\geq1.
\end{eqnarray*}

Thus, cluster expansions for the sequence of probability distribution functions \eqref{cexp} of a system
of particles interacting with their nearest neighbors have a structure determined by mapping \eqref{cl},
namely,
\begin{eqnarray*}
   &&\hskip-5mm I+D(t)=(I-g(t))^{-{\mathbb I}_{\star}}.
\end{eqnarray*}
The structure of cumulant expansions for the sequence of correlation functions \eqref{cuexp} is defined
by the inverse mapping \eqref{cu}
\begin{eqnarray*}
   &&\hskip-5mm g(t)=I-(I+D(t))^{-{\mathbb I}_{\star}}.
\end{eqnarray*}

The origin of the structure of expansions \eqref{cexp} and \eqref{сcuexp} will be argued in Section 5.
We note that in the case of many particles whose state is described by symmetric distribution functions,
the structure of cluster and cumulant expansions is determined by exponential and logarithmic mappings
defined by the suitable tensor $\ast$-product \cite{GG22}, respectively.

We will also introduce some generalizations of the concept of cumulant expansions. Namely, we will define cumulant
expansions for a cluster of particles $(-s_1,\ldots,s_2)$ and particles $(-(n_{1}+s_{1}),\ldots,-(s_{1}+1)),
(s_{2}+1,\ldots,s_{2}+n_{2})$, i.e. correlation functions of a cluster of particles and particles
\begin{eqnarray}\label{сcuexp}
   &&\hskip-12mm g_{1+n_1+n_2}(x_{-(n_{1}+s_{1})},\ldots,x_{-(s_{1}+1)},\{x_{-s_1},\ldots,x_{s_2}\},
       x_{s_{2}+1},\ldots,x_{s_{2}+n_{2}})=\\
    && \sum_{\mbox{\scriptsize$\begin{array}{c}\mathrm{P}:(x_{-(n_{1}+s_{1})},\ldots,x_{-(s_{1}+1)},\{x_{-s_1},\ldots,x_{s_2}\},\\
        x_{s_{2}+1},\ldots,x_{s_{2}+n_{2}})=\bigcup_i X_i\end{array}$}}(-1)^{|\mathrm{P}|-1}
        \prod_{X_i\subset \mathrm{P}}D_{|\theta(X_i)|}(\theta(X_i)),\quad n_{1}+n_{2}\geq 0.\nonumber
\end{eqnarray}
In the expansion for the $(1+n_1+n_2)$-th order cumulant (semi-invariants) of probability distribution
functions, the following notation is used: the symbol $\sum_\mathrm{P}$ denotes the sum over each ordered
partition of $\mathrm{P}$ of the partially ordered set of indices $(-(n_{1}+s_{1}),\ldots,-(s_{1}+1),
\{-s_1,\ldots,s_2\},s_{2}+1,\ldots,s_{2}+n_{2})$ into $|\mathrm{P}|$ nonempty partially ordered
subsets $X_i$ that do not intersect each other, the connected set $\{-s_1,\ldots,s_2\}$ belongs
entirely to one of the subsets $X_i$, and the declusterization mapping $\theta(\cdot)$ is defined
by the equality: $\theta(\{X_i\})=(X_i)$.

We emphasize that cumulants (\ref{сcuexp}) are solutions to the recursion relations known as the cluster
expansions of the groups of operators (\ref{S*}) of a cluster of particles and particles, namely,
\begin{eqnarray}\label{сcexp}
   &&\hskip-12mm D_{n_1+s_1+s_2+n_2}(x_{-(n_{1}+s_{1})},\ldots,x_{-(s_{1}+1)},x_{-s_1},\ldots,x_{s_2},
       x_{s_{2}+1},\ldots,x_{s_{2}+n_{2}})=\\
   &&\sum\limits_{\mbox{\scriptsize $\begin{array}{c}\mathrm{P}:(x_{-(n_{1}+s_{1})},\ldots,
       x_{-(s_{1}+1)},\{x_{-s_1},\ldots,x_{s_2}\},\\ x_{s_{2}+1},\ldots,x_{s_{2}+n_{2}}))
       \bigcup_{i}X_{i}\end{array}$}}\prod_{X_i\subset \mathrm{P}}g_{|X_i|}(t,X_i), \quad n_1+n_2\geq1.\nonumber
\end{eqnarray}

The correlation functions of a particle cluster and particles are related to the correlation functions
of particles by the following relations:
\begin{eqnarray}\label{rel}
  &&\hskip-12mm g_{1+n_{1}+n_{2}}(t,x_{-(n_{1}+s_{1})},\ldots,x_{-(s_{1}+1)},\{x_{-s_1},\ldots,x_{s_2}\},
         x_{s_{2}+1},\ldots,x_{s_{2}+n_{2}})=\\
  &&\hskip-7mm  \sum_{\mbox{\scriptsize$\begin{array}{c}\mathrm{P}:(x_{-(n_{1}+s_{1})},\ldots,
       x_{-(s_{1}+1)},\{x_{-s_1},\ldots,x_{s_2}\},\\
       x_{s_{2}+1},\ldots,x_{s_{2}+n_{2}})=\bigcup_i X_i\end{array}$}}
      (-1)^{|\mathrm{P}|-1}\prod_{X_i\subset \mathrm{P}}\,
      \sum\limits_{\mathrm{P'}:\,\theta(X_{i})=\bigcup_{j_i} Z_{j_i}}
      \prod_{Z_{j_i}\subset \mathrm{P'}}g_{|Z_{j_i}|}(t,Z_{j_i}).\nonumber
\end{eqnarray}
Let us provide instances of these relations:
\begin{eqnarray*}
   &&\hskip-12mm g_{1+0+0}(\{x_1\})=g_{0+1}(x_1), \\
   &&\hskip-12mm g_{1+0+0}(\{x_{-1},x_1\})=g_{1+1}(x_{-1},x_1)+g_{1+0}(x_{-1})g_{0+1}(x_1),\\
   &&\hskip-12mm g_{1+0+1}(\{x_{-1},x_1\},x_2)=g_{1+2}(x_{-1},x_1,x_2)+g_{1+1}(x_{-1},x_1)g_{0+1}(x_2)+\\
   &&g_{1+0}(x_{-1})g_{0+2}(x_1,x_2)+g_{1+0}(x_{-1})g_{0+1}(x_1)g_{0+1}(x_2),\\
   &&\vdots
\end{eqnarray*}
In the particular case $n_{1}+n_{2}=0$, i.e., a cluster consisting of $s_1+s_2$ particles, relations
\eqref{rel} take the form
\begin{eqnarray*}
  &&g_{1+0+0}(t,\{x_{-s_1},\ldots,x_{s_2}\})=
      \sum\limits_{\mathrm{P}:\,\theta(\{x_{-s_1},\ldots,x_{s_2}\})=\bigcup_{i} X_{i}}
      \prod_{X_{i}\subset\mathrm{P}}g_{|X_{i}|}(t,X_{i}).\nonumber
\end{eqnarray*}

Note that the structure of cluster and cumulant expansions for systems of particles with topological
interaction of nearest neighbors is determined by the symmetry of the functions to the renumbering of
their phase variables and is not related to the dimensionality of the space, in which the particles move.

\textcolor{blue!50!black}{\section{A hierarchy of evolution equations for cumulants of distribution functions}}
As noted above, the sequence of correlation functions \eqref{cuexp} describes the state of systems of a non-fixed
number of particles in an alternative way to the sequence of probability distribution functions \cite{PG83}.

The evolution of the state of a system of particles with topological interaction is described in terms of
correlation functions which are governed by the Liouville hierarchy of the nonlinear equations:
\begin{eqnarray}\label{Lh}
   &&\hskip-12mm\frac{\partial}{\partial t}g_{s_1+s_2}(t,x_{-s_1},\ldots,x_{s_2})=
      \mathcal{L}^*_{s_1+s_2}g_{s_1+s_2}(t,x_{-s_1},\ldots,x_{s_2})+\\
   &&\sum\limits_{\mathrm{P}:\,(x_{-s_1},\ldots,x_{s_2})=X_{1}\bigcup X_2}\,
      \sum\limits_{j=\mathrm{max}\widehat{X}_{1}}
      \mathcal{L}_{\mathrm{int}}^{\ast}(j,j+1)g_{|X_{1}|}(t,X_{1})g_{|X_{2}|}(t,X_{2}),\nonumber \\
      \nonumber \\
  \label{Lhi}
   &&\hskip-12mm g_{s_1+s_2}(t)\big|_{t=0}=g_{s_1+s_2}^{0},\quad s_1+s_2\geq1,
\end{eqnarray}
where the Liouville operator $\mathcal{L}^*_{s_1+s_2}$ is defined by expression (\ref{Lstar}),
${\sum\limits}_{\mathrm{P}:\,(x_{-s_1},\ldots,x_{s_2})=X_{1}\bigcup X_2}$ is the sum over all possible ordered
partitions of $\mathrm{P}$ of the partially ordered set $(x_{-s_1},\ldots,x_{s_2})$ into two subsets $X_1$ and
$X_2$ that do not intersect, the symbol $\widehat{X}_i$ denotes the set of indices of the corresponding set
$X_i$ of phase coordinates of particles, and $\mathrm{max}\widehat{X}_{1}$ is the maximum value of the set of
indices $\widehat{X}_{1}$.

We emphasize that the Liouville hierarchy (\ref{Lh}) is a hierarchy of recurrent evolution equations.
We give examples of evolution equations \eqref{Lh}:
\begin{eqnarray*}
   &&\hskip-8mm\frac{\partial}{\partial t}g_{0+1}(t,x_1)=\mathcal{L}^\ast(1)g_{1}(t,x_1),\\
   &&\hskip-8mm\frac{\partial}{\partial t}g_{1+1}(t,x_{-1},x_1)=\big(\sum_{j\in(-1,1)}\mathcal{L}^{\ast}(j)+
       \mathcal{L}_{\mathrm{int}}^\ast(-1,1)\big)g_{1+1}(t,x_{-1},x_1)+\\
   &&\hskip+8mm \mathcal{L}_{\mathrm{int}}^\ast(-1,1)g_{1+0}(t,x_{-1})g_{0+1}(t,x_{1}),\\
   &&\hskip-8mm\frac{\partial}{\partial t}g_{1+2}(t,x_{-1},x_1,x_2)=\big(\sum_{j\in(-1,1,2)}\mathcal{L}^{\ast}(j)+
       \sum_{(j,j+1)\in ((-1,1),(1,2))}\mathcal{L}^{\ast}_{\mathrm{int}}(j,j+1)\big)g_{1+2}(t,x_{-1},x_1,x_2)+\\
   &&\hskip+8mm \mathcal{L}_{\mathrm{int}}^\ast(-1,1)g_{1+0}(t,x_{-1})g_{0+2}(t,x_{1},x_2)+
        \mathcal{L}_{\mathrm{int}}^\ast(1,2)g_{1+1}(t,x_{-1},x_1)g_{0+1}(t,x_2), \\
   &&\vdots
\end{eqnarray*}

To construct a solution to the Cauchy problem of the Liouville hierarchy (\ref{Lh}), we introduce the concept
of cumulants of the groups of operators of the Liouville equations, i.e., a connected part of the groups
of operators of the Liouville equations. The cumulant of the $(s_1+s_2)$-th order of the groups of operators
\eqref{S*} is defined by the following expansion
\begin{eqnarray}\label{cum}
   &&\hskip-12mm \mathfrak{A}_{s_1+s_2}(t,-s_1,\ldots,s_2)\doteq
      \sum\limits_{\mbox{\scriptsize $\begin{array}{c}\mathrm{P}:(-s_1,\ldots,s_2)=
       \bigcup_{i}X_{i}\end{array}$}}(-1)^{|\mathrm{P}|-1}\,\prod_{X_i\subset \mathrm{P}}S^*_{|X_i|}(t,X_i),
\end{eqnarray}
where the symbol $\sum_\mathrm{P}$ denotes the sum over all ordered partitions $\mathrm{P}$ of the partially
ordered set of indexes $(-s_1,\ldots,-1,1,\ldots,s_2)$ into $|\mathrm{P}|$ nonempty partially ordered
subsets $X_i\in(-s_1,\ldots,-1,1,\ldots,s_2)$ that do not intersect. Here are some examples of cumulant
expansions \eqref{cum}:
\begin{eqnarray*}
   &&\hskip-12mm \mathfrak{A}_{0+1}(t,1)=S^*_{0+1}(t,1), \\
   &&\hskip-12mm \mathfrak{A}_{1+1}(t,-1,1)=S^*_{1+1}(t,-1,1)-S^*_{0+1}(t,-1)S^*_{0+1}(t,1),\\
   &&\hskip-12mm \mathfrak{A}_{1+2}(t,-1,1,2)=S^*_{1+2}(t,-1,1,2)-S^*_{1+1}(t,-1,1)S^*_{0+1}(t,2)-\\
   &&S^*_{1+0}(t,-1)S^*_{0+2}(t,1,2)+S^*_{1+0}(t,-1)S^*_{0+1}(t,1)S^*_{0+1}(t,2),\\
   &&\vdots
\end{eqnarray*}

We remark that cumulants \eqref{cum} are solutions of the recursive equations known as cluster
expansions of the groups of operators \eqref{S*} of the Liouville equations
\begin{eqnarray}\label{exp}
   &&\hskip-12mm S^*_{s_1+s_2}(t,-s_1,\ldots,s_2)=
      \sum\limits_{\mbox{\scriptsize $\begin{array}{c}\mathrm{P}:(-s_1,\ldots,s_2)=
       \bigcup_{i}X_{i}\end{array}$}}\prod_{X_i\subset \mathrm{P}}\mathfrak{A}_{|X_i|}(t,X_i),
       \quad s_1+s_2\geq1.
\end{eqnarray}
The structure of these cluster expansions is determined by the structure of the generator \eqref{Lstar} of
the group of operators \eqref{S*} for topologically interacting particles.

The evolution of initial cumulants of distribution functions \eqref{сcuexp} is described by the following
correlation functions:
\begin{eqnarray}\label{sLhc}
   &&\hskip-7mm g_{s_1+s_2}(t,x_{-s_1},\ldots,x_{s_2})=
       \sum\limits_{\mathrm{P}:\,(x_{-s_1},\ldots,x_{s_2})=\bigcup_j X_j}
      \prod_{X_j\subset \mathrm{P}}\mathfrak{A}_{|\widehat{X}_j|}(t,\widehat{X}_j)
      g^{0}_{|\mathrm{P}|}(\{X_1\},\ldots,\{X_{|\mathrm{P}|}\}),\\
   &&\hskip-7mm s_1+s_2\geq1,\nonumber
\end{eqnarray}
where the symbol $\sum_\mathrm{P}$ denotes the sum over all ordered partitions $\mathrm{P}$ of the partially
ordered set of indexes $(x_{-s_1},\ldots,x_{s_2})$ into $|\mathrm{P}|$ nonempty partially ordered subsets
$X_i\in(x_{-s_1},\ldots,x_{s_2})$ that do not intersect, the symbol $\widehat{X}_i$ denotes the set of
indices of the corresponding set $X_i$ of phase coordinates of particles, and the generating operators of
these expansions are products of cumulants \eqref{cum} of the groups of operators \eqref{S*}. We adduce some
examples of correlation functions \eqref{sLhc}
\begin{eqnarray*}
   &&g_{0+1}(t,x_1)=\mathfrak{A}_{1}(t,1)g_{0+1}^0(x_1),\\
   &&g_{1+1}(t,x_{-1},x_1)=\mathfrak{A}_{2}(t,-1,1)g_{1+0}^{0}(\{x_{-1},x_{1}\})+
     \mathfrak{A}_{1}(t,-1)\mathfrak{A}_{1}(t,1)g_{1+1}^{0}(x_{-1},x_{1}),\\
   &&g_{3}(t,x_{-1},x_1,x_2)=\mathfrak{A}_{3}(t,-1,1,2)g_{1+0}^{0}(\{x_{-1},x_1,x_2\})+
      \mathfrak{A}_{2}(t,-1,1)\mathfrak{A}_{1}(t,2)g_{1+1}^{0}(\{x_{-1},x_1\},x_2)+\\
   &&\hskip+7mm+\mathfrak{A}_{1}(t,-1)\mathfrak{A}_{2}(t,1,2)g_{1+1}^{0}(x_{-1},\{x_1,x_2\})+
      \mathfrak{A}_{1}(t,-1)\mathfrak{A}_{1}(t,1)\mathfrak{A}_{1}(t,2))g_{1+2}^{0}(x_{-1},x_1,x_2),\\
   &&\vdots
\end{eqnarray*}

Indeed, by applying cluster expansions \eqref{exp} of the groups of operators \eqref{S*} to operator
groups in the definition of cumulants \eqref{cuexp} of distribution functions \eqref{D} and considering
the definition of initial correlation functions \eqref{cuexp}, we derive expansions \eqref{sLhc}.

For arbitrary initial states from the space of double sequences of integrable functions the solution
of the Cauchy problem of the Liouville hierarchy \eqref{Lh},\eqref{Lhi} is also represented by the
following expansions which are equivalent to expansions \eqref{sLhc}:
\begin{eqnarray}\label{sLh}
    &&\hskip-7mm g_{s_1+s_2}(t,x_{-s_1},\ldots,x_{s_2})=
        \sum\limits_{\mathrm{P}:\,(x_{-s_1},\ldots,x_{s_2})=\bigcup_i X_i}
        \mathfrak{A}_{|\mathrm{P}|}(t,\{\widehat{X}_1\},\ldots,\{\widehat{X}_{|\mathrm{P}|}\})
        \prod_{X_i\subset \mathrm{P}}g_{|X_i|}^0(X_i),\\
   &&\hskip-7mm s_1+s_2\geq1.\nonumber
\end{eqnarray}
The generating operator $\mathfrak{A}_{|\mathrm{P}|}(t)$ of the expansion \eqref{sLh} is the $|\mathrm{P}|$-th
order cumulant \eqref{cum} of the groups of operators \eqref{S*} which is represented by the following expansion
\begin{eqnarray}\label{cumulantP}
   &&\hskip-12mm \mathfrak{A}_{|\mathrm{P}|}(t,\{\widehat{X}_1\},\ldots,\{\widehat{X}_{|\mathrm{P}|}\})\doteq
      \sum\limits_{\mathrm{P}^{'}:\,(\{\widehat{X}_1\},\ldots,\{\widehat{X}_{|\mathrm{P}|}\})=
      \bigcup_k \widehat{Z}_k}(-1)^{|\mathrm{P}^{'}|-1}
      \prod\limits_{\widehat{Z}_k\subset\mathrm{P}^{'}}S^*_{|\theta(\widehat{Z}_{k})|}(t,\theta(\widehat{Z}_{k})),
\end{eqnarray}
where the connected set of ordered indices $\widehat{X}_i$, i.e., one element of the set, is denoted by the symbol
$\{\widehat{X}_i\}$, the declusterization mapping $\theta(\cdot)$ is defined by the equality:
$\theta(\{\widehat{X}_i\})=(\widehat{X}_i)$ and the symbol
$\sum_{\mathrm{P}^{'}:\,(\{\widehat{X}_1\},\ldots,\{\widehat{X}_{|\mathrm{P}|}\})=\bigcup_k \widehat{Z}_k}$ denotes
the sum over all possible ordered partitions $\mathrm{P}^{'}$ of the ordered set of indices
$(\{\widehat{X}_1\},\ldots,\{\widehat{X}_{|\mathrm{P}|}\})$ on $|\mathrm{P}^{'}|$ nonempty partially ordered subsets
$\widehat{Z}_k\subset (\{\widehat{X}_1\},\ldots,$ $\{\widehat{X}_{|\mathrm{P}|}\})$.
We adduce some examples of correlation functions \eqref{sLh}
\begin{eqnarray*}
    &&\hskip-7mm g_{0+1}(t,x_1)=\mathfrak{A}_{1}(t,1)g_{0+1}^0(x_1),\\
    &&\hskip-7mm g_{1+1}(t,x_{-1},x_1)=\mathfrak{A}_{1}(t,\{-1,1\})g_{1+1}^0(x_{-1},x_1)+
       \mathfrak{A}_{2}(t,-1,1)g_{1+0}^0(x_{-1})g_{0+1}^0(x_1),\\
    &&\hskip-7mm g_{1+2}(t,x_{-1},x_1,x_2)=\mathfrak{A}_{1}(t,\{-1,1,2\})g_{1+2}^0(x_{-1},x_1,x_2)+
       \mathfrak{A}_{2}(t,-1,\{1,2\})g_{1+0}^0(x_{-1})g_{0+2}^0(x_1,x_2)+\\
    &&\mathfrak{A}_{2}(t,\{-1,1\},2)g_{0+1}^0(x_2)g_{1+1}^0(x_{-1},x_1)+
       \mathfrak{A}_{3}(t,-1,1,2)g_{1+0}^0(x_{-1})g_{0+1}^0(x_1)g_{0+1}^0(x_2),\\
   &&\vdots
\end{eqnarray*}
The validity of the expansions \eqref{sLh} for the solution of the Liouville hierarchy
\eqref{Lh} is proved by point-wise differentiation for the time variable.

The following existence theorem for the Liouville hierarchy (\ref{Lh}) holds.

\smallskip
\begin{theorem}
For $t\in\mathbb{R}$ the solution of the Cauchy problem of the Liouville hierarchy
\eqref{Lh},\eqref{Lhi} is represented by expansions \eqref{sLh}. For infinitely differentiable
initial correlation functions with compact supports from the space $L^{1}_{n_1+n_2},\,n_1+n_2\geq1$,
expansions \eqref{sLh} represent the strong (classical) solution and for arbitrary initial correlation
functions from the space $L^{1}_{n_1+n_2},\,n_1+n_2\geq1$ it is the weak (generalized) solution.
\end{theorem}

We emphasize that the characteristic properties of the constructed solution \eqref{sLh} are generated
by the properties of the cumulants \eqref{cumulantP} of the groups of operators (\ref{S*}) of the
Liouville equations.

The following criterion holds.

\smallskip
\begin{criterion}
A solution of the Cauchy problem for the Liouville hierarchy \eqref{Lh},\eqref{Lhi} for particle systems
with topological interaction is represented by expansions \eqref{sLh} if and only if the generating operators
of expansions \eqref{sLh} are solutions of cluster expansions (\ref{exp}) of the groups of operators (\ref{S*}).
\end{criterion}

The necessary condition means that cluster expansions (\ref{exp}) are valid for groups of operators (\ref{S*}).
These recurrence relations are derived from the definition (\ref{cuexp}) of correlation functions, provided
that they are represented as expansions \eqref{sLh} for the solution of the Cauchy problem to the Liouville
hierarchy \eqref{Lh},\eqref{Lhi}.

The sufficient condition means that the infinitesimal generator of one-parameter mapping \eqref{sLh} coincides
with the generator of the Liouville hierarchy (\ref{Lh}), which is the consequence of Theorem 1.

\textcolor{blue!50!black}{\section{The BBGKY hierarchy for topologically interacting particles}}
To describe the evolution of states of both finite and infinite numbers of particles, an alternative
approach equivalent to the one presented above is used. This approach is based on the description of
the evolution of states by reduced distribution functions.

Further, we will formulate an approach to describing the evolution of the state of a system
of many particles with topological interaction in terms of reduced distribution functions defined
within the framework of the dynamics of correlations \eqref{Lh}. Recall \cite{PG83} that reduced
distribution functions are defined through the sequence
$D(0)=\{D^0_{n_1+n_2}(x_{-n_1},\ldots,x_{-1},x_{1},\ldots,x_{n_2})\}_{n_1+n_2 \geq 0}$
of probability distribution functions at the initial instant and the groups of operators (\ref{S*}):
\begin{eqnarray}\label{ms}
     &&\hskip-12mm F_{s_1+s_2}(t,x_{-s_1},\ldots,x_{s_2})\doteq(I,D(0))^{-1}\sum_{n=0}^{\infty}
       \sum_{\mbox{\scriptsize$\begin{array}{c}n=n_{1}+n_{2}\\n_{1},n_{2}\geq 0\end{array}$}}
       \int_{(\mathbb{R}\times\mathbb{R})^{n_{1}+n_{2}}}dx_{-(n_{1}+s_{1})}\ldots\\
    &&\hskip-7mm dx_{-(s_{1}+1)}dx_{s_{2}+1}\ldots dx_{s_{2}+n_{2}}\,
      (S^*(t)D(0))_{n_{1}+s_1+s_2+n_{2}}(x_{-(n_{1}+s_{1})},\ldots,x_{s_{2}+n_{2}}),
      \quad s_{1}+s_{2}\geq 1,\nonumber
\end{eqnarray}
where the normalizing factor
\begin{eqnarray*}
  &&\hskip-12mm (I,D(0))\doteq
      \sum_{n=0}^{\infty}\sum_{\mbox{\scriptsize$\begin{array}{c}n=n_{1}+n_{2}\\n_{1},n_{2}\geq 0\end{array}$}}
      \int_{(\mathbb{R}\times\mathbb{R})^{n_{1}+n_{2}}}dx_{-n_{1}}\ldots dx_{n_{2}}
      D^0_{n_1+n_2}(x_{-n_1},\ldots,x_{-1},x_{1},\ldots,x_{n_2})
\end{eqnarray*}
known as the grand canonical partition function. The justification for introducing this definition of reduced
distribution functions for topologically nearest-neighbor interacting particles is provided in \cite{PG83}.
Notice that the sequence of reduced distribution functions \eqref{ms} satisfies the BBGKY hierarchy \eqref{BBGKY}
for topologically nearest-neighbor interacting particles \cite{G1}.

The result of dividing the series in expression \eqref{ms} by the series of the normalizing factor naturally
allows one to redefine the reduced distribution functions as a series expansion over the cumulants of
distribution functions \eqref{cuexp}
\begin{eqnarray*}
    &&\hskip-12mm F_{s_1+s_2}(t,x_{-s_1},\ldots,x_{s_2})\doteq
      \sum_{n=0}^{\infty}\sum_{\mbox{\scriptsize $\begin{array}{c}n=n_{1}+n_{2}\\n_{1},n_{2}\geq 0\end{array}$}}
      \int_{(\mathbb{R}\times\mathbb{R})^{n_{1}+n_{2}}}dx_{-(n_{1}+s_{1})}\ldots
      dx_{-(s_{1}+1)}dx_{s_{2}+1}\ldots \\
   &&\hskip-7mm dx_{s_{2}+n_{2}}\sum_{\mbox{\scriptsize$\begin{array}{c}\mathrm{P}:(x_{-(n_{1}+s_{1})},\ldots,
       x_{-(s_{1}+1)},\{x_{-s_1},\ldots,x_{s_2}\},\\
       x_{s_{2}+1},\ldots,x_{s_{2}+n_{2}})=\bigcup_i X_i\end{array}$}}
       (-1)^{|\mathrm{P}|-1}\,\prod_{X_i\subset \mathrm{P}}(S^*(t)D(0))_{|X_i|}(X_i),\\
   &&\hskip-12mm s_{1}+s_{2}\geq 1.
\end{eqnarray*}
Thus, the definition of reduced distribution functions, equivalent to the definition of \eqref{ms}, is
formulated based on correlation functions (\ref{cuexp}) by the following series expansions:
\begin{eqnarray}\label{FClusters}
    &&\hskip-12mm F_{s_1+s_2}(t,x_{-s_1},\ldots,x_{s_2})\doteq
      \sum_{n=0}^{\infty}\sum_{\mbox{\scriptsize $\begin{array}{c}n=n_{1}+n_{2}\\n_{1},n_{2}\geq 0\end{array}$}}
       \int_{(\mathbb{R}\times\mathbb{R})^{n_{1}+n_{2}}}dx_{-(n_{1}+s_{1})}\ldots dx_{-(s_{1}+1)}dx_{s_{2}+1}\ldots \\
   &&\hskip-7mm dx_{s_{2}+n_{2}}\,
      g_{1+n_{1}+n_{2}}(t,x_{-(n_{1}+s_{1})},\ldots,x_{-(s_{1}+1)},\{x_{-s_1},\ldots,x_{s_2}\},
      x_{s_{2}+1},\ldots,x_{s_{2}+n_{2}}),\nonumber\\
   &&\hskip-12mm s_{1}+s_{2}\geq 1.\nonumber
\end{eqnarray}
On the set of allowed configurations, the correlation functions of the particle cluster
and particles $g_{1+n_{1}+n_{2}}(t),\,n_{1}+n_{2}\geq 1,$ are defined by the following expansions:
\begin{eqnarray}\label{rozLh}
    &&\hskip-12mm g_{1+n_{1}+n_{2}}(t,x_{-(n_{1}+s_{1})},\ldots,x_{-(s_{1}+1)},\{x_{-s_1},\ldots,x_{s_2}\},
         x_{s_{2}+1},\ldots,x_{s_{2}+n_{2}})=\\
    &&\hskip-5mm \sum_{\mbox{\scriptsize$\begin{array}{c}\mathrm{P}:(x_{-(n_{1}+s_{1})},\ldots,
       x_{-(s_{1}+1)},\{x_{-s_1},\ldots,x_{s_2}\},\\
       x_{s_{2}+1},\ldots,x_{s_{2}+n_{2}})=\bigcup_i X_i\end{array}$}}
       \mathfrak{A}_{|\mathrm{P}|}\big(t,\{\theta(\widehat{X}_1)\},\ldots,
       \{\theta(\widehat{X}_{|\mathrm{P}|})\}\big)
       \prod_{X_i\subset \mathrm{P}}g_{|X_i|}^0(X_i).\nonumber
\end{eqnarray}
Recall that the generating operators of the expansions \eqref{rozLh} are represented by the
$|\mathrm{P}|th$-th order cumulants \eqref{cumulantP} of the groups of operators \eqref{S*} and
the symbol $\sum_{\mathrm{P}}$ denotes the sum over all ordered partitions of the set of indices
$(-(n_{1}+s_{1}),\ldots,-(s_{1}+1),\{-s_1,\ldots,-1,1,\ldots,s_2\},s_{2}+1,\ldots,s_{2}+n_{2})$
into intersecting subsets $X_i$.

Thus, we observe that the structure of cumulant expansions for states \eqref{cexp} for topologically
interacting particles is determined by the method incorporating the normalizing factor in the mean
value functional which is determined through reduced distribution functions.

Let's introduce the space $L_\alpha^1=\sum_{n=0}^{\infty}\sum_{\mbox{\scriptsize$\begin{array}{c}n=n_{1}+n_{2}\\
n_{1},n_{2}\geq 0\end{array}$}}\bigoplus\alpha^{n_1+n_2} L_{n_1+n_2}^1$ of double sequences
$f=\{f_{n_1+n_2}\}_{n_1+n_2\geq0},$ of integrable functions $f_{n_1+n_2}(x_{-n_1},\ldots,x_{n_2})$,
defined on the phase space
$\mathbb{R}^{n_1+n_2}\times(\mathbb{R}^{n_1+n_2}\setminus\mathbb{W}_{n_1+n_2})$ with the norm
\begin{eqnarray*}
   &&\|f\|=\sum_{n=0}^{\infty}\sum_{\mbox{\scriptsize$\begin{array}{c}n=n_{1}+n_{2}\\
       n_{1},n_{2}\geq 0\end{array}$}}\alpha^{n_1+n_2}\int_{(\mathbb{R}\times\mathbb{R})^{n_{1}+n_{2}}}
       dx_{-n_1}\ldots dx_{n_2}\,|f_{n_1+n_2}(x_{-n_1},\ldots,x_{n_2})|,
\end{eqnarray*}
where the parameter $\alpha>1$.

The following bounded operators are defined on sequences of functions $f\in L_\alpha^1$:
\begin{eqnarray}\label{op}
   &&\big(\mathfrak{a}_{+}f\big)_{n_1+n_2}(x_{-n_1},\ldots,x_{n_2})=
      \int_{\mathbb{R}\times\mathbb{R}}dx_{n_2+1}\,f_{n_1+n_2+1}(x_{-n_1},\ldots,x_{n_2},x_{n_2+1}),\\
   &&\big(\mathfrak{a}_{-}f\big)_{n_1+n_2}(x_{-n_1},\ldots,x_{n_2})=
      \int_{\mathbb{R}\times\mathbb{R}}dx_{-(n_1+1)}\,f_{n_1+1+n_2}(x_{-(n_1+1)},x_{-n_1},\ldots,x_{n_2}),\nonumber
\end{eqnarray}
which are analogs of the particle annihilation operator of quantum field theory, as well as the following
operators:
\begin{eqnarray}\label{opp}
   &&(I-\mathfrak{a}_{\pm})^{-1}=\sum_{n=0}^{\infty}\mathfrak{a}_{\pm}^n.
\end{eqnarray}

Since the correlation functions \eqref{rozLh} are determined by the Liouville hierarchy \eqref{Lh},
the reduced distribution functions \eqref{FClusters} satisfy the BBGKY hierarchy for a system of many
topologically interacting particles \cite{G1}
\begin{eqnarray}\label{BBGKY}
   &&\frac{d}{dt}F(t)=\mathcal{L}^*F(t)+[\mathfrak{a}_{-},\mathcal{L}^*]F(t)+
                     [\mathfrak{a}_{+},\mathcal{L}^*]F(t),
\end{eqnarray}
where $\mathcal{L}^*=\sum_{n=0}^{\infty}\sum_{\mbox{\scriptsize$\begin{array}{c}n=n_{1}+n_{2}\\
n_{1},n_{2}\geq 0\end{array}$}}\bigoplus\mathcal{L}^*_{n_1+n_2}$ is the direct sum of the Liouville
operators \eqref{Lstar}, the brackets $[ \circ , \circ ]$ denote the commutator of two operators, and,
therefore, in component form, these BBGKY generator expressions are represented as follows:
\begin{eqnarray*}
   &&\hskip-9mm \big([\mathfrak{a}_{-},\mathcal{L}^*]F(t)\big)_{n_1+n_2}(x_{-n_1},\ldots,x_{n_2})=
          \int_{\mathbb{R}\times\mathbb{R}} dx_{-(n_1+1)}\,\frac{\partial}{\partial q_{-n_1}}\Phi(q_{-n_1}-q_{-(n_1+1)})
          \frac{\partial}{\partial p_{-n_1}}\,F_{n_1+1+n_2}(t),\\
   &&\hskip-9mm \big([\mathfrak{a}_{+},\mathcal{L}^*]F(t)\big)_{n_1+n_2}(x_{-n_1},\ldots,x_{n_2})=
         \int_{\mathbb{R}\times\mathbb{R}} dx_{n_2+1}\,\frac{\partial }{\partial q_{n_2}}\Phi(q_{n_2}-q_{n_2+1})
          \frac{\partial}{\partial p_{n_2}}\,F_{n_1+n_2+1}(t).
\end{eqnarray*}

A solution of the Cauchy problem to the BBGKY hierarchy \eqref{BBGKY} of a many-particle system
with topological nearest-neighbor interaction is represented by such series expansions
for the sequence of reduced distribution functions \cite{GerS,GerRS,G92}:
\begin{eqnarray}\label{Bhsol}
   &&\hskip-10mm F_{s_1+s_2}(t,x_{-s_1},\ldots,x_{s_2})=
      \sum_{n=0}^{\infty}\sum_{\mbox{\scriptsize $\begin{array}{c}n=n_{1}+n_{2}\\n_{1},n_{2}\geq 0\end{array}$}}
       \int_{(\mathbb{R}\times\mathbb{R})^{n_{1}+n_{2}}}dx_{-(n_{1}+s_{1})}\ldots dx_{-(s_{1}+1)}\times \\
   &&\hskip-5mm dx_{s_{2}+1}\ldots dx_{s_{2}+n_{2}}\,\mathfrak{A}_{1+n_1+n_2}(t,-(n_{1}+s_{1}),
       \ldots,-(s_{1}+1),\{-s_1,\ldots,s_2\},\nonumber\\
   &&\hskip-5mm s_{2}+1,\ldots,s_{2}+n_{2} )\,
        F_{s_1+n_1+s_2+n_2}^0(x_{-(n_{1}+s_{1})},\ldots,x_{s_{2}+n_{2}}),\quad s_{1}+s_{2}\geq 1,\nonumber
\end{eqnarray}
where the generating operators of these series are cumulants of the groups of operators \eqref{S*} of
the Liouville equations
\begin{eqnarray}\label{Bhcum}
   &&\hskip-12mm \mathfrak{A}_{1+n_1+n_2}(t)=
       \sum_{\mbox{\scriptsize$\begin{array}{c}\mathrm{P}:(-(n_{1}+s_{1}),\ldots,-(s_{1}+1),\{-s_1,\ldots,s_2\},\\
        s_{2}+1,\ldots,s_{2}+n_{2})=\bigcup_i X_i\end{array}$}}(-1)^{|\mathrm{P}|-1}
        \prod_{X_i\subset \mathrm{P}}S^*_{|\theta(X_i)|}(t,\theta(X_i)).
\end{eqnarray}
In the expansion for the $(1+n_1+n_2)$-th order cumulant of the groups of operators (\ref{S*}),
the following notation is used: the symbol $\sum_\mathrm{P}$ denotes the sum over each ordered
partition of $\mathrm{P}$ of the partially ordered set of indices $(-(n_{1}+s_{1}),\ldots,-(s_{1}+1),
\{-s_1,\ldots,s_2\},s_{2}+1,\ldots,s_{2}+n_{2})$ into $|\mathrm{P}|$ nonempty partially ordered
subsets $X_i$ that do not intersect each other, and the connected set $\{-s_1,\ldots,s_2\}$ belongs
entirely to one of the subsets $X_i$.

We emphasize that cumulants (\ref{Bhcum}) are solutions to the recursion relations known as the cluster
expansions \eqref{cum} of the groups of operators (\ref{S*}), namely
\begin{eqnarray}\label{expc}
   &&\hskip-12mm S^*_{n_1+s_1+s_2+n_2}(t,-(n_{1}+s_{1}),\ldots,-(s_{1}+1),-s_1,\ldots,s_2,
        s_{2}+1,\ldots,s_{2}+n_{2}))=\\
   &&\sum\limits_{\mbox{\scriptsize $\begin{array}{c}\mathrm{P}:(-(n_{1}+s_{1}),\ldots,-(s_{1}+1),\{-s_1,\ldots,s_2\},\\
        s_{2}+1,\ldots,s_{2}+n_{2})
       \bigcup_{i}X_{i}\end{array}$}}\prod_{X_i\subset \mathrm{P}}\mathfrak{A}_{|X_i|}(t,X_i),
       \quad n_1+n_2\geq1.\nonumber
\end{eqnarray}

In terms of the introduced operators \eqref{opp} in the space $L_\alpha^1$, the series expansions
\eqref{Bhsol} of the sequence of reduced distribution functions have the following structure
\begin{eqnarray}\label{Bhs}
   &&F(t)=(I-\mathfrak{a}_{-})^{-1}(I-\mathfrak{a}_{+})^{-1}(I-S^*(t))^{-{\mathbb I}_{\star}}F(0).
\end{eqnarray}

For the initial reduced distribution functions from the space $L_\alpha^1$ series \eqref{Bhsol}
is defined. Since for the generating operators \eqref{Bhcum} of series expansions \eqref{Bhsol}
in the space $L_{s_1+n_1+s_2+n_2}^1$ the following estimate holds:
\begin{eqnarray*}
 &&\|\mathfrak{A}_{1+n_1+n_2}(t)\,f_{s_1+n_1+s_2+n_2}\|_{L_{{s_1+n_1+s_2+n_2}}^1}\leq
    2^{n_1+n_2}\,\|f_{s_1+n_1+s_2+n_2}\|_{L_{{s_1+n_1+s_2+n_2}}^1},
\end{eqnarray*}
then for the sequence of functions \eqref{Bhsol} in the space $L_\alpha^1$, under the condition
$\alpha>2,$ we have the following estimate
\begin{eqnarray*}
 &&\|F(t)\|_{L_{\alpha}^1}\leq c_\alpha\|F^0\|_{L_{\alpha}^1},
\end{eqnarray*}
where $c_\alpha =(1-\frac{2}{\alpha})^{-1}$. The parameter $\alpha$ can be interpreted as a quantity
inversely proportional to the density of a system of many particles, that is, the average number of
particles per unit volume.

Thus, for a non-perturbative solution \eqref{Bhsol} of the Cauchy problem for the BBGKY hierarchy
of particle systems with topological interaction, the following theorem holds \cite{GerS},\cite{GerRS}.

\smallskip
\begin{theorem}
If $F(0)\in L_{\alpha}^1$ is a double sequence of non-negative initial distribution functions, then
for $t\in\mathbb{R}$ under the condition $\alpha >2,$ there exists a unique solution of the Cauchy
problem for the BBGKY hierarchy \eqref{BBGKY} for particle systems with topological interaction,
which is represented by series expansions \eqref{Bhsol}.
This is a strong solution for initial data from the subspace of finite sequences of continuously
differentiable functions with compact supports. For arbitrary initial data from the space $L_{\alpha}^1$,
it is a weak solution.
\end{theorem}

The following criterion holds.

\smallskip
\begin{criterion}
A solution of the Cauchy problem for the BBGKY hierarchy \eqref{BBGKY} for particle systems with topological
interaction is represented by series expansions \eqref{Bhsol} if and only if the generating operators of series
expansions \eqref{Bhsol} are solutions of cluster expansions (\ref{expc}) of the groups of operators (\ref{S*}).
\end{criterion}

The necessary condition means that cluster expansions (\ref{expc}) are valid for groups of operators (\ref{S*}).
These recurrence relations are derived from the definition (\ref{ms}) of reduced distribution functions, provided
that they are represented as series expansions \eqref{Bhsol} for the solution of the Cauchy problem to the BBGKY
hierarchy (\ref{BBGKY}).
The sufficient condition means that the infinitesimal generator of one-parameter mapping \eqref{Bhs} coincides
with the generator of the BBGKY hierarchy (\ref{BBGKY}), which is the consequence of Theorem 2.

In the articles \cite{G1},\cite{PG83}, a solution of the Cauchy problem for the BBGKY hierarchy
\eqref{BBGKY} was constructed in a reduced form, the expression of which in terms of operators
\eqref{opp} has the following structure
\begin{eqnarray*}
   &&F(t)=(I-\mathfrak{a}_{-})^{-1}(I-\mathfrak{a}_{+})^{-1}S^*(t)(I-\mathfrak{a}_{-})(I-\mathfrak{a}_{+})F(0).
\end{eqnarray*}
Component-wise, this sequence of reduced distribution functions is represented by such series expansions:
\begin{eqnarray}\label{rs}
     &&\hskip-12mm F_{s_{1}+s_{2}}(t,x_{-s_{1}},\ldots,x_{s_{2}})=
       \sum_{n=0}^{\infty}\sum_{\mbox{\scriptsize $\begin{array}{c}n=n_{1}+n_{2}\\n_{1},n_{2}\geq 0\end{array}$}}
       \int_{(\mathbb{R}\times\mathbb{R})^{n_{1}+n_{2}}}dx_{-(n_{1}+s_{1})}\ldots dx_{-(s_{1}+1)}\times \\
    &&\hskip-5mm dx_{s_{2}+1}\ldots dx_{s_{2}+n_{2}}\,
       U_{1+n_{1}+n_{2}}(t,-(n_{1}+s_{1}),\ldots,-(s_1+1),\{-s_1\ldots,-1,1,\ldots,s_2\},\nonumber\\
    &&\hskip-5mm s_2+1,\ldots,s_{2}+n_{2})F_{n_{1}+s_{1}+s_{2}+n_{2}}^0(x_{-(n_{1}+s_{1})},\ldots,x_{s_{2}+n_{2}}),
    \quad s_{1}+s_{2} \geq 1.\nonumber
\end{eqnarray}
were the generating operators $\{U_{1+n_{1}+n_{2}}(t)\}_{n_1+n_2\geq0},$ of series \eqref{rs} are determined
by the following expansions:
\begin{eqnarray*}
    &&\hskip-12mm U_{1+n_{1}+n_{2}}(t,-(n_{1}+s_{1}),\ldots,-(s_1+1),\{-s_1,\ldots,-1,1,\ldots,s_2\},
       s_2+1,\ldots,s_{2}+n_{2})=\\
    &&\hskip-7mm \sum_{k_{1}=0}^{\textrm{min}(1,n_1)}\sum_{k_{2}=0}^{\textrm{min}(1,n_2)}(-1)^{k_1+k_2}\,
       S^*_{n_{1}+s_{1}-k_1+s_{2}+n_{2}-k_2}(t,-(n_{1}+s_{1}-k_{1}),\ldots,-1,1,\ldots,s_{2}+n_{2}-k_{2}).
\end{eqnarray*}
Such expressions are solutions of reduced cluster expansions of the groups of operators \eqref{S*}
in terms of cumulants (semi-invariants).

In the space $L_{\alpha}^1$ of sequences of integrable functions, due to the isometrics of the
group of operators \eqref{S*}, the representations \eqref{Bhsol} and \eqref{rs} of the solution
of the Cauchy problem for the hierarchy of BBGKY equations \eqref{BBGKY} are equivalent. We
emphasize that a traditional representation of the solution of the Cauchy problem for the BBGKY
hierarchy is the series expansion of the perturbation theory \cite{CGP97},\cite{G92}. Under
appropriate conditions for the initial data and the interaction potential of particles, the
equivalence of the representations of these solutions is proved due to the validity of the
analogs of the Duhamel equations for the generating operators of the series \eqref{rs}.

\textcolor{blue!50!black}{\section{Some observations on reduced correlation functions}}
The characteristics of fluctuations of the mean value of observables on a macroscopic scale
are directly determined by reduced correlation functions describing the state of a system of
many particles on a microscopic scale \cite{GG21}.

Then an alternative approach to describing the evolution of topologically interacting
particles is based on reduced correlation functions defined by the  cumulant expansions over
the reduced distribution functions \eqref{ms}:
\begin{eqnarray}\label{G}
   &&\hskip-12mm G_{s_1+s_2}(t,x_{-s_1},\ldots,x_{s_2})=
      \sum\limits_{\mbox{\scriptsize $\begin{array}{c}\mathrm{P}:(x_{-s_1},\ldots,x_{s_2})=
       \bigcup_{i}X_{i}\end{array}$}}(-1)^{|\mathrm{P}|-1}\prod_{X_i\subset\mathrm{P}}F_{|X_i|}(t,X_i),
       \quad s_1+s_2\geq1,
\end{eqnarray}
or in terms of sequences of functions, these expansions are defined by the inverse mapping \eqref{cu}
and take the form
\begin{eqnarray*}
   &&\hskip-5mm G(t)=I-(I+F(t))^{-{\mathbb I}_{\star}},
\end{eqnarray*}
where the used denotations are similar to that of the recursive equations \eqref{cexp}.

Then the definition of reduced correlation functions is formulated based on the solution (\ref{sLh})
of the Cauchy problem for the hierarchy of the Liouville equations \eqref{Lh},\eqref{Lhi} by the
following expansions into series:
\begin{eqnarray}\label{Gexpg}
   &&\hskip-12mm G_{s_1+s_2}(t,x_{-s_1},\ldots,x_{s_2})\doteq
      \sum_{n=0}^{\infty}\sum_{\mbox{\scriptsize $\begin{array}{c}n=n_{1}+n_{2}\\n_{1},n_{2}\geq 0\end{array}$}}
      \int_{(\mathbb{R}\times\mathbb{R})^{n_{1}+n_{2}}}dx_{-(n_{1}+s_{1})}\ldots dx_{-(s_{1}+1)}\times \\
   &&dx_{s_{2}+1}\ldots dx_{s_{2}+n_{2}}\,g_{n_{1}+s_{1}+s_{2}+n_{2}}(t,x_{-(n_{1}+s_{1})},\ldots,x_{s_{2}+n_{2}}),
   \quad s_{1}+s_{2} \geq 1,\nonumber
\end{eqnarray}
where the generating functions $g_{n_{1}+s_{1}+s_{2}+n_{2}}(t)$ are determined by the expansions \eqref{sLh}.
In terms of sequences of functions, the sequence of reduced correlation functions \eqref{Gexpg} has the following
structure
\begin{eqnarray*}
   &&G(t)=(I-\mathfrak{a}_{-})^{-1}(I-\mathfrak{a}_{+})^{-1}g(t).
\end{eqnarray*}

We emphasize that the $(n_{1}+n_{2})$-th term of expansions \eqref{Gexpg} of reduced correlation
functions is determined by the correlation function of $(n_{1}+s_{1}+s_{2}+n_{2})$ particles \eqref{sLh},
in contrast to the expansions of reduced distribution functions \eqref{FClusters}, which are determined
by the $(1+n_{1}+n_{2})$-th correlation function of a particle cluster and particles \eqref{rozLh}.

Since the correlation functions $g_{n_{1}+s_{1}+s_{2}+n_{2}}(t)$ are determined by the hierarchy of the
Liouville equations (\ref{Lh}), the reduced correlation functions \eqref{Gexpg} satisfy the hierarchy
of nonlinear evolution equations.

Note that for many-particle systems whose states are described by symmetric distribution functions,
a non-perturbative solution of the Cauchy problem to the BBGKY hierarchy was constructed in the works
\cite{GG22},\cite{GerR,GG23,GG24}.

\textcolor{blue!50!black}{\section{Conclusion}}
The article establishes that the dynamics of correlations, i.e., the fundamental evolution equations
\eqref{Lh}, form the basis for describing the evolution of all possible states of both a finite and
an infinite number of particles with topological interaction \eqref{Lstar}. This approach is based
on describing the state using functions determined by cluster expansions of probability distribution
functions \eqref{cexp}. Let us emphasize that the correlations created during the evolution of a system
of particles with topological interaction \eqref{rozLh} by nature differ from the structure of correlations
of particle systems whose state is traditionally described by symmetric distribution functions \cite{GG22}.

It was also proved above that the constructed correlation dynamics forms the basis for describing the
evolution of the state of infinite particle systems in terms of reduced distribution functions \eqref{FClusters}
or reduced correlation functions \eqref{Gexpg}, namely, cumulants of reduced distribution functions \eqref{G}.

The structure of expansions for correlation functions \eqref{sLh}, the generating operators of which
are the cumulants \eqref{cumulantP} of the groups of operators \eqref{S*} of the corresponding order,
induces the cumulant structure of series expansions for reduced distribution functions \eqref{Bhsol}
and reduced correlation functions, respectively. Thus, the dynamics of systems of an infinite number
of particles with topological nearest-neighbor interaction are generated by the dynamics of correlations
of particle states \eqref{Lh}.

In conclusion, we note that the statistical properties of physical systems are modeled by the collective
behavior of infinite particle systems \cite{CGP97}. It should also note the importance of a mathematical
description of the processes of creation and propagation of correlations, in particular, for numerous
applications. In the case of systems of colliding particles, the dynamics of correlations were constructed
in \cite{GG22}, and for quantum many-particle systems in \cite{G12,G17,G21}.

\vskip+7mm

\noindent \textbf{Acknowledgements.} \,\,{\large\textcolor{blue!55!black}{Glory to Ukra\"{\i}ne!}}

\bigskip

\addcontentsline{toc}{section}{\textcolor{blue!55!black}{References}}

\vskip+5mm

\appendix

\textcolor{blue!55!black}{\section{Appendix: The hierarchy of evolution equations for observables}}

In view of the mean value functional of observables determines the duality of observables and the
state of particle systems, we will consider another approach to describing the evolution of a system
of a non-fixed number of ordered particles with the topological interaction of nearest neighbors.
This equivalent approach is based on the concept of observables governed by the hierarchy of evolution
equations for reduced observables.

In the case of a non-fixed number of many identical ordered particles interacting with their nearest
neighbors through a potential defined above, the mean value of the double sequences of reduced observables
$B_{n_1+n_2}(t),\,n_1+n_2\geq0,$ is determined by a continuous linear functional represented by the following
expansion into the series \cite{PG83}
\begin{eqnarray*}
   &&\hskip-7mm\big(B(t),F(0)\big)\doteq\\
   &&\sum_{s=0}^{\infty}\sum_{{\scriptsize \begin{array}{c} s=s_{1}+s_{2}\\
       s_{1},s_{2}\geq 0\end{array}}}
       \int_{\left(\mathbb{R}\times\mathbb{R}\right)^{s_1+s_2}}
        dx_{-s_1}\ldots dx_{s_2}B_{s_{1}+s_{2}}(t,x_{-s_{1}},\ldots,x_{s_{2}})
        F_{s_{1}+s_{2}}^0(x_{-s_{1}},\ldots,x_{s_{2}}).
\end{eqnarray*}

Let $C_{n_1+n_2}\equiv C(\mathbb{R}^{n_1+n_2}\times(\mathbb{R}^{n_1+n_2}\setminus\mathbb{W}_{n_1+n_2}))$
be the space of bounded continuous functions which are equal to zero on the set of forbidden configurations
$\mathbb{W}_{n_1+n_2}$. The following operators are defined on sequences of functions
$b=\{b_{n_1+n_2}(x_{-n_1},\ldots,x_{n_2})\}_{n_1+n_2\geq0}$ from the space $b_{n_1+n_2}\in C_{n_1+n_2}$:
\begin{eqnarray}\label{ac}
   &&\big(\mathfrak{a}^{+}_{+}b \big)_{n_1+n_2}(x_{-n_1},\ldots,x_{n_2})=b_{n_1+n_2-1}(x_{-n_1},\ldots,x_{n_2-1}),\\
   &&\big(\mathfrak{a}^{+}_{-}b \big)_{n_1+n_2}(x_{-n_2},\ldots,x_{n_1})=b_{n_1-1+n_2}(x_{-(n_1-1)},\ldots,x_{n_2}),\nonumber
\end{eqnarray}
which are analogs of the particle creation operator of quantum field theory, as well as the operators and
their inverses:
\begin{eqnarray*}
   &&\big(I-\mathfrak{a}^{+}_{\pm}\big)^{-1}=\sum_{n=0}^{\infty}(\mathfrak{a}^{+}_{\pm})^n, \quad
     \big(I-\mathfrak{a}^{+}_{\pm}\big)=I-\mathfrak{a}^{+}_{\pm}.
\end{eqnarray*}
In terms of operators \eqref{opp} the mean value functional takes the form
\begin{eqnarray}\label{mvo}
   &&\big(B(t),F(0)\big)=\big((I-\mathfrak{a}_{-})^{-1}(I-\mathfrak{a}_{+})^{-1}B(t)F(0)\big)_0.
\end{eqnarray}
Note that operators \eqref{ac} are adjoint to operators \eqref{op} in the sense of functional \eqref{mvo},
i.e., $\big(\mathfrak{a}^{+}_{\pm}\,b,f\big)=\big(b,\mathfrak{a}_{\pm}f\big)$.

On sequences of functions $b=\{b_{n_1+n_2}(x_{-n_1},\ldots,x_{n_2})\}_{n_1+n_2\geq0}$ from the space
$C_{n_1+n_2}$, the direct sum of the adjoint groups of operators to the operators \eqref{S*} is defined,
$\big(S(t)\,b,f\big)=\big(b,S^\ast(t)f\big)$, and, consequently,
\begin{eqnarray}\label{S}
   &&S_{n_1+n_2}(t)=S^\ast_{n_1+n_2}(-t).
\end{eqnarray}
Accordingly, on finite sequences of differential functions, a direct sum of the Liouville generators adjoint
to the operators \eqref{Lstar} is defined, $\big(\mathcal{L}\,b,f\big)=\big(b,\mathcal{L}^\ast f\big)$,
and, consequently, $\mathcal{L}_{n_1+n_2}=-\mathcal{L}^\ast_{n_1+n_2}$.

Recall \cite{PG83} that a sequence of reduced observables are defined using the sequence of initial
observables $A(0)=\{A^0_{n_1+n_2}(x_{-n_1},\ldots,x_{-1},x_{1},\ldots,x_{n_2})\}_{n_1+n_2 \geq 0}$
and the groups of operators (\ref{S}):
\begin{eqnarray}\label{dro}
   &&B(t)=(I-\mathfrak{a}^{+}_{-})(I-\mathfrak{a}^{+}_{+})S(t)A(0),
\end{eqnarray}
or in a component-wise form reduced observables are represented by the following expansions:
\begin{eqnarray*}
   &&\hskip-7mm  B_{s_1+s_2}(t,x_{-s_{2}},\ldots,x_{s_{1}})=
         ((I-\mathfrak{a}^{+}_{-})(I-\mathfrak{a}^{+}_{+})S(t)A(0))_{s_1+s_2}(x_{-s_{2}},\ldots,x_{s_{1}})=\\
   &&\sum_{n_{1}=0}^{\textrm{min}(1,s_{1})}\sum_{n_{2}=0}^{\textrm{min}(1,s_{2})}(-1)^{n_1+n_2}\,
       S_{s_1-n_1+s_2-n_2}(t)A_{s_1-n_1+s_2-n_2}^0(x_{-(s_{1}-n_{1})},\ldots,x_{s_{2}-n_{2}}).
\end{eqnarray*}

The sequence of reduced observables (\ref{dro}) satisfies the hierarchy of evolution equations for reduced
observables \cite{CGP97}. In term of operators (\ref{ac}) this hierarchy has the following abstract form
\begin{eqnarray*}
   &&\frac{\partial}{\partial t}B(t)=\big(I-\mathfrak{a}^{+}_{-}\big)\big(I-\mathfrak{a}^{+}_{+}\big)\mathcal{L}
                    \big(I-\mathfrak{a}^{+}_{-}\big)^{-1}\big(I-\mathfrak{a}^{+}_{+}\big)^{-1}B(t),
\end{eqnarray*}
or in the case of a pair interaction potential under consideration
\begin{eqnarray}\label{dBBGKY}
   &&\frac{\partial}{\partial t}B(t)=\mathcal{L}B(t)+[\mathcal{L},\mathfrak{a}^{+}_{-}]B(t)+
                     [\mathcal{L},\mathfrak{a}^{+}_{+}]B(t),
\end{eqnarray}
where the brackets $[\circ,\circ]$ denote the commutator of two operators.

Therefore, in a component-wise form, the Cauchy problem for the hierarchy of evolution equations (\ref{dBBGKY})
for reduced observables is represented as follows:
\begin{eqnarray*}
   &&\hskip-5mm\frac{\partial}{\partial t}B_{s_1+s_2}(t,x_{-s_1},\ldots,x_{s_2})=
                        \mathcal{L}_{s_1+s_2}(-s_1,\ldots,s_2)B_{s_1+s_2}(t,x_{-s_1},\ldots,x_{s_2})+\\
   &&+\mathcal{L}_{\mathrm{int}}(-s_{1},-s_{1}+1)B_{s_1-1+s_2}(t,x_{-s_{1}+1},\ldots,x_{s_2})
     +\mathcal{L}_{\mathrm{int}}(s_{2}-1,s_{2})B_{s_1+s_2-1}(t,x_{-s_1},\ldots,x_{s_2-1}),\\ \\
   &&\hskip-5mm B_{s_1+s_2}(t,x_{-s_1},\ldots,x_{s_2})= B_{s_1+s_2}^0(x_{-s_1},\ldots,x_{s_2}), \quad s_1+s_2\geq 1.
\end{eqnarray*}


A solution to the Cauchy problem for the hierarchy of reduced observables in a component-wise form is represented
by the expansions:
\begin{eqnarray}\label{sro}
   &&\hskip-12mm B_{s_1+s_2}(t,x_{-s_1},\ldots,x_{s_2})=\sum_{n_{1}=0}^{s_{1}}\sum_{n_{2}=0}^{s_{2}}\,\,
       \mathfrak{A}_{1+s_1-n_1+s_2-n_2}(t,-s_{1},\ldots,-(n_{1}+1),\\
   &&\hskip+9mm\{-n_{1},\ldots,-1,1,\ldots,n_{2}\},n_{2}+1,\ldots,s_2)
                B_{{n_1+n_2}}^0(x_{-n_{1}},\ldots,x_{-1},x_{1},\ldots,x_{n_2}),\nonumber \\
   &&\hskip-12mm s_1+s_2\geq 1.\nonumber
\end{eqnarray}
The generating operators of these expansions are corresponding-order cumulants of the group
of operators (\ref{S})
\begin{eqnarray}\label{cumo}
   &&\hskip-12mm \mathfrak{A}_{1+s_1-n_1+s_2-n_2}(t,
        -s_{1},\ldots,-(n_{1}+1),\{-n_{1},\ldots,-1,1,\ldots,n_{2}\},n_{2}+1,\ldots,s_2)=\\
   &&\hskip-7mm \sum_{\mbox{\scriptsize$\begin{array}{c}\mathrm{P}:(
        -s_{1},\ldots,-(n_{1}+1),\{-n_{1},\ldots,-1,1,\ldots,n_{2}\},\\ n_{2}+1,\ldots,s_2)=\bigcup_i X_i\end{array}$}}
      (-1)^{|\mathrm{P}|-1}\prod_{X_i\subset \mathrm{P}}S_{|\theta(X_i)|}(t,\theta(X_i)),\nonumber
\end{eqnarray}
where the notations adopted in formula \eqref{сcuexp} are used.
We observe that cumulants (\ref{cumo}) are solutions to the recursion relations known as the cluster
expansions of the groups of operators (\ref{S}):
\begin{eqnarray}\label{cexo}
  &&\hskip-7mm S_{s_1+s_2}(t,-s_1,\ldots,-(n_{1}+1),-n_{1},\ldots,-1,1,\ldots,n_{2},n_{2}+1,\ldots,s_2)=\\
  &&\sum\limits_{\mbox{\scriptsize $\begin{array}{c}\mathrm{P}:(-s_1,\ldots,-(n_{1}+1),\{-n_{1},\ldots,-1,
        1,\ldots,n_{2}\},\\ n_{2}+1,\ldots,s_2)=
       \bigcup_{i}X_{i}\end{array}$}}\prod_{X_i\subset \mathrm{P}}\mathfrak{A}_{|X_i|}(t,X_i),\nonumber
\end{eqnarray}

In terms of the introduced above operators, the series expansions representing a solution of the Cauchy
problem to the hierarchy of evolution equations for reduced observables has the following structure
\begin{eqnarray}\label{asro}
   &&B(t)=\big(I-\mathfrak{a}^{+}_{-}\big)\big(I-\mathfrak{a}^{+}_{+}\big)\big(I-(I+S(t))^{-\mathbb{I}_{\star}}\big)B(0).
\end{eqnarray}

The following criterion holds.

\smallskip
\begin{criterion}
A solution of the Cauchy problem of the hierarchy of reduced observables (\ref{dBBGKY}) is represented
by expansions (\ref{sro}) if and only if the generating operators of expansions (\ref{sro} are solutions
of cluster expansions (\ref{cexo}) of the groups of operators (\ref{S}).
\end{criterion}

The necessary condition means that cluster expansions (\ref{cexo}) are valid for groups of operators
(\ref{S}). These recurrence relations are derived from definition (\ref{dro}) of reduced
observables, provided that they are represented as expansions (\ref{sro}) for the solution
of the Cauchy problem of the hierarchy of evolution equations for reduced observables (\ref{dBBGKY}).

The sufficient condition means that the infinitesimal generator of one-parameter mapping (\ref{asro})
coincides with the generator of the hierarchy of evolution equations for reduced observables (\ref{dBBGKY}).
\begin{theorem}
A non-perturbative solution of the Cauchy problem of the hierarchy of evolution equations for reduced
observables (\ref{dBBGKY}) is represented by expansions (\ref{sro}) in which the generating
operators are cumulants of the corresponding order (\ref{cumo}) of the groups of operators (\ref{S}).
For initial data $B(0)\in C_{0}$ of finite sequences of infinitely differentiable functions with
compact supports sequence (\ref{sro}) is a unique global-in-time classical solution and for
arbitrary initial data $B(0)\in C_{\gamma}$ is a unique global-in-time generalized solution.
\end{theorem}

Note that cluster expansions \eqref{cexo} of the groups of operators \eqref{S} underlie the
classification of possible non-perturbative solution representations of the Cauchy problem
of the hierarchy of evolution equations for reduced observables (\ref{dBBGKY}). For example,
solutions of the recursive relations \eqref{cexo} to first-order cumulants can be represented
as expansions in terms of cumulants acting on variables on which initial reduced observables
depend, and in terms of cumulants not acting on these variables. Then the component-wise sequence
of reduced observables (\ref{sro}) is represented by such expansions into series:
\begin{eqnarray*}
   &&\hskip-12mm B_{s_1+s_2}(t,x_{-s_1},\ldots,x_{s_2})=\sum_{n_{1}=0}^{s_{1}}\sum_{n_{2}=0}^{s_{2}}\,
       \sum_{k_{1}=0}^{\textrm{min}(1,s_{1}-n_1)}\sum_{k_{2}=0}^{\textrm{min}(1,s_{2}-n_2)}(-1)^{k_1+k_2}\,
       S_{s_1-k_1+s_2-k_2}(t,\\
   &&-(s_{1}-k_{1}),\ldots,-1,1,\ldots,s_{2}-k_{2})B_{{n_1+n_2}}^0(x_{-n_{1}},\ldots,x_{-1},x_{1},\ldots,x_{n_2}),
       \quad s_1+s_2\geq 1.\nonumber
\end{eqnarray*}
In terms of operators \eqref{ac}, the series expansions of the sequence of reduced observables have
the following structure:
\begin{eqnarray*}
   &&B(t)=(I-\mathfrak{a}^{+}_{-})(I-\mathfrak{a}^{+}_{+})S(t)(I-\mathfrak{a}^{+}_{-})^{-1}(I-\mathfrak{a}^{+}_{+})^{-1}B(0).
\end{eqnarray*}

We note that in the case of many-particle systems whose observables are symmetric functions,
the structure of corresponding cluster and cumulant expansions was analyzed in \cite{GG21},\cite{GG23}.

\end{document}